\documentclass[aps, prb, twocolumn, superscriptaddress, footinbib]{revtex4-2}

\usepackage[x11names,table]{xcolor}

\usepackage[]{hyperref}
\hypersetup{
  colorlinks,
  linkcolor={IndianRed3},
  citecolor={Cyan4},
  urlcolor={Cyan4}
}

\usepackage{graphicx}

\newcommand{\subfigref}[2]{\ref{#1}\hyperref[#1]{\textbf{(#2)}}}

\usepackage{mathtools, physics,bbm}

\newcommand{\NESS}{non-equilibrium steady state}
\newcommand{\qd}{QD}
\newcommand{\res}{QMR}
\newcommand{\mm}{self-oscillations}
\newcommand{\dnt}{torotropy}
\newcommand{\TT}{\mathcal{T}}
\newcommand{\id}{\mathbbm{1}}
\newcommand{\av}[1]{\langle #1 \rangle}
\newcommand{\R}{\mathrm{leads}}
\renewcommand{\d}{\hat{d}}
\renewcommand{\b}{\hat{b}}
\newcommand{\bb}{\hat{b}^\dagger}
\renewcommand{\c}{\hat{c}_{\nu}}

\newcommand{\I}{I_\nu^\mathrm{el}}
\newcommand{\IL}{I_L^\mathrm{el}}
\newcommand{\IR}{I_R^\mathrm{el}}
\newcommand{\trho}{\tilde{\rho}}
\newcommand{\JL}{\tilde{J}_L}
\newcommand{\JR}{\tilde{J}_R}
\newcommand{\Power}{\tilde{P}^\mathrm{el}_\mathrm{tot}}
\newcommand{\J}{\tilde{J}_\nu}
\newcommand{\G}{\Upsilon_{\nu}}
\newcommand{\Rnn}[6]{\text{A}^{{#1}{#2}}_{{#3}{#4}\,{#5}{#6}}\Big|_\nu}
\newcommand{\Qnn}[6]{Q^{{#1}\rightarrow {#2}}_{{#3}{#4}\,{#5}{#6}}\Big|_\nu}
\newcommand{\D}{\hat{D}}
\newcommand{\DD}{\hat{D}^\dagger}
\newcommand{\e}[2]{\omega_{{#1},{#2}}}
\let\Tr\realx
\DeclareMathOperator{\Tr}{Tr}

\begin{document}

\title{Autonomous conversion of particle-exchange to quantum self-oscillations}

\author{Sofia Sevitz}
\email{sevitz@uni-potsdam.de}
\affiliation{University of Potsdam, Institute of Physics and Astronomy, Karl-Liebknecht-Str. 24-25, 14476 Potsdam, Germany\looseness=-1}

\author{Federico Cerisola}
\affiliation{Physics and Astronomy, University of Exeter, Exeter EX4 4QL, United Kingdom}

\author{Karen V. Hovhannisyan}
\affiliation{University of Potsdam, Institute of Physics and Astronomy, Karl-Liebknecht-Str. 24-25, 14476 Potsdam, Germany\looseness=-1}

\author{Janet Anders}
\affiliation{University of Potsdam, Institute of Physics and Astronomy, Karl-Liebknecht-Str. 24-25, 14476 Potsdam, Germany\looseness=-1}
\affiliation{Physics and Astronomy, University of Exeter, Exeter EX4 4QL, United Kingdom}

\date{\today} 

\begin{abstract}

Particle-exchange machines utilize electronic transport to continuously transfer heat between fermionic reservoirs. Here, we couple a quantum mechanical resonator to a particle-exchange machine hosted in a quantum dot and let the system run autonomously. This way, part of the energy exchanged between the reservoirs can be stored in the resonator in the form of self-oscillations. Our analysis goes well beyond previous works by exploring the slow transport regime and accessing arbitrarily strong dot--resonator coupling. First, we introduce a faithful measure of self-oscillations, and use it to certify that they can occur in the slow-transport regime. We furthermore show that the electrical current through the dot can be used to witness self-oscillations. Finally, we establish that, under realistic conditions, self-oscillations occur only when the machine operates as a heater. We define an experimentally measurable performance metric characterizing the efficiency of current--to--self-oscillations conversion. It reveals that, counterintuitively, strong dot--resonator coupling is detrimental to the conversion performance. The framework developed here can be readily implemented in a variety of nanoscale devices, such as a suspended carbon nanotube with an embedded quantum dot. 

\end{abstract}

\maketitle

\section{Introduction}

Interest has grown towards platforms that allow the construction of
particle-exchange (PE) machines~\cite{Humphrey2005, Ludovico2016, Esposito2009, Muralidharan2012, Manikandan2020}. Their operation is based on exchanging electrons among fermionic reservoirs (or leads) driven by a chemical and/or temperature imbalance~\cite{Humphrey2002, Scovil1959, Topp2015}. They are said to have \textit{continuous strokes} due to the fact that the leads and system are always in contact, thus transferring energy continuously~\cite{Uzdin2015}. This is why they are typically studied in the (non-equilibrium) steady state regime. Contrary to cyclical engines, they do not require moving parts controlled by an external drive, and are therefore classified as \textit{autonomous machines}~\cite{MarinGuzman2024, Benenti_2017, Niedenzu2019}.

Previous works predicted that by using an energy filter between the fermionic reservoirs, a PE heat engine can be constructed with excellent thermoelectric conversion efficiency, even saturating the Carnot bound~\cite{Humphrey2002, Mahan1996, Esposito2009, Nakpathomkun2010, Borga2012, Esposito2012}. The first experimental verification employing a semiconductor quantum dot (\qd) as the energy filter was reported in Ref.~\cite{Josefsson2018}. There, they built a heat engine whose efficiency reached 70\% of the Carnot efficiency while still depositing finite power output into an external resistor, a purely dissipative load. Such a choice of load is typically informed by the macroscopic intuition that a non-dissipative work-like load would not alter the operation of the machine~\cite{Leijnse2010, Whitney_2014, Benenti_2017}. However, when we move to the nanoscale quantum regime, this assumption does not, in general, hold~\cite{Aberg_2013, Horodecki_2013, Gilz_2013,Mcconnell2022}.

An important next step is to understand how PE machines responds to a quantum work-like load. To that end, here we study a nanoscale platform that can host PE while allowing for the power output to be stored in a separate, quantum, degree of freedom. In this way, the produced energy can be either directly measured or stored, as in a battery, to power another device. Such a device can be of much interest to a multitude of scientific fields spanning from thermodynamics~\cite{Roche2015, Mari2015, Verteletsky2020, Campaioli_2024} to information~\cite{Strasberg2013, Koski2015, Sanchez2019}.

\begin{figure}[t]
  \centering
  \includegraphics[width=1\linewidth]{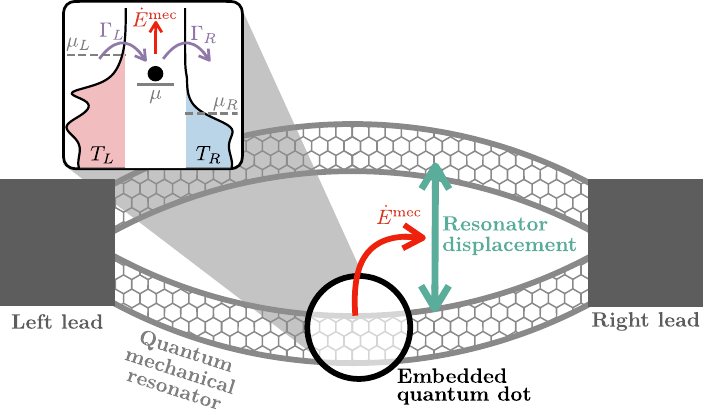}
  \caption{
  \textbf{Particle exchange to mechanical motion machine.} 
  A quantum dot (inset) with chemical potential $\mu$ is coupled to two fermionic leads to the left ($L$) and right ($R$) that hold a temperature $T_L\geq T_R$ and/or chemical $\mu_L \geq \mu_R$ imbalance, and have arbitrary spectral densities (black wiggly lines). The imbalance enables hopping of electrons from one reservoir to the other with electronic tunneling rates $\Gamma_{L/R}$ exchanging electrical energy and, due to the coupling between the dot occupation and the mechanical displacement, produces mechanical energy flow $\dot{E}^\mathrm{mec}$ that is stored in the displacement of the quantum mechanical resonator, thus acting as a work load. 
  }
  \label{fig:Fig1}
\end{figure}

Here we study a promising platform for this purpose: a \qd~embedded in a quantum mechanical resonator (\res), see Fig.~\ref{fig:Fig1}. The \qd~serves as the medium of the PE machine whose produced energy can then be stored in the displacement of the \res, acting as a work load. The resulting mechanical motion is in the form of \mm~as they are induced internally by the coupling between the \res~and the \qd~\cite{Jenkins2013, Culhane2022, Wachtler2019}. This model can be realized experimentally, for example, with a \qd~embedded in a suspended carbon nanotube (CNT)~\cite{Huttel2009, Luo2017, Meerwaldt2012, Steele2009, Vigneau2022, Wen2020}. These devices are very promising for nanoscale thermodynamics as they can possess very large quality factors~\cite{Laird2012, Moser2014}. Most theoretical work has focused on the fast tunneling (also known as quasi-adiabatic) regime, where the natural frequency of the resonator is slow compared to the electronic tunneling rates. In such a case, a semi-classical approach is valid~\cite{Wen2020, Monsel2020}, and the standard approaches of classical thermodynamics can be applied~\cite{Wachtler2019, Culhane2022, De2016, Tabanera2024}. However, given the potential of these devices to reach very high quality factors with high resonance frequencies~\cite{Laird2012} and at low temperatures, they provide a very promising platform to reach the quantum regime for the mechanical degree of freedom~\cite{Samanta2023}. This opens the doors for the explorations of thermodynamic concepts in the quantum domain~\cite{Potts2024, Vinjanampathy2016}.

\section{Model description}\label{Sec:PlatformDescription}

The total Hamiltonian is composed of the system part $\hat{H}_s$ (\qd~+ \res) coupled via $\hat{V}$ to two fermionic reservoirs which we denote with the subscript ``$\R$'',
\begin{equation}\label{eq:H_total}
  \hat{H}= \hat{H}_s+\hat{V}+\hat{H}_\R,
\end{equation}
see Fig.~\ref{fig:Fig1} for reference. Throughout the article we take the natural units $\hbar=k_B=1$. Each component of Eq.~\eqref{eq:H_total} reads as~\cite{Culhane2022, De2016, Mitra2004, Leijnse2010}
\begin{align}\label{eq:H_s}
  \hat{H}_s &= \mu \d^\dagger \d + \omega \bb \b + \omega \lambda (\bb+\b)\d^\dagger \d
  \\ \label{eq:H_R}
  \hat{H}_\R &=\sum_{\nu} \epsilon_{\nu} \c^\dagger \c, \\ \label{eq:V_r}
  \hat{V} &=\sum_{\nu} t_\nu \c^\dagger \d + \mathrm{h.c.} .
\end{align}

Eq.~\eqref{eq:H_s} constitutes the system Hamiltonian which is composed of a two-level \qd~with chemical potential $\mu$ coupled to a \res~with a single mechanical frequency $\omega$. On the one hand, $\d^\dagger$ is the electronic creation operator of the \qd. Thus $\hat{n} = \d^\dagger \d$ is the electronic population of the dot where $\langle \hat{n} \rangle$ can only take the values from $0$ to $1$; for empty and occupied, respectively. On the other hand, $\bb$ is the phonon creation operator acting on the \res. The third term of Eq.~\eqref{eq:H_s} describes the coupling between the electronic population of the dot and the displacement of the resonator governed by the parameter $\lambda$. Here we treat this coupling non-perturbatively, thus capturing the back-action of the \qd~onto the \res, and vice versa.

A lead to the left ($L$) and another to the right ($R$) of the \qd~(see inset of Fig.~\ref{fig:Fig1}) are described by Eq.~\eqref{eq:H_R}, where $\epsilon_{\nu}$ is the free energy of a spinless electron in the reservoir $\nu=\{L,R\}$ and $\c^\dagger$ is its creation operator. The leads are large and always in thermal equilibrium, with their fermionic occupations dictated by their respective Fermi functions as $f_\nu(E)=1/(\exp(\beta_\nu(E-\mu_\nu))+1)$, with $\mu_\nu$ the chemical potential and $\beta_\nu=1/T_\nu$ inverse temperature. The leads can exchange particles with the \qd~via the dot-reservoir coupling detailed in Eq.~\eqref{eq:V_r}, where $t_\nu$ are the coupling constants determined by the tunneling rates $\G(\epsilon)=2\pi |t_\nu|^2\delta(\epsilon-\epsilon_{\nu})$. We consider these rates to be centered around $\gamma_\nu$ with broadening $\delta_\nu$ modeled by a single Lorentzian shape $\G(\epsilon) = \Gamma_\nu \delta_\nu^2 / \big[(\epsilon-\gamma_\nu)^2 +\delta_\nu^2\big],$ with $\Gamma_\nu$ the dot-reservoir tunneling rate~\cite{Erpenbeck2018_2,Schaller2013}. 

We explicitly do not couple the \res~to an external bath. This is done so that any effect observed on the \res~is solely due to energy transferred from the PE. In a real device there is also additional damping of the \res~oscillations by its own coupling to a heat bath.

\medskip

To study the operation of the autonomous machine, we consider the slow tunneling regime, where single electron sequential tunneling is valid, i.e.
\begin{equation}\label{eq:Conditions}
  \Gamma_\nu \ll T_\nu, \omega .
\end{equation}
To proceed we use the recently derived master equation~\cite{Sevitz2025} to compute the \NESS~of the system $\rho_s=\Tr_\R(\rho)$, where $\rho$ is the density matrix associated with $\hat{H}$ (defined in Eq.~\eqref{eq:H_total}). In Appendix~\ref{Appendix:Master_equation} we summarize the main equations of Ref.~\cite{Sevitz2025} used in this paper.

In the following section we explore under which conditions the \res~utilizes the energy injected by the particle transport to generate mechanical motion in the form of \mm. After faithfully identifying \mm, the autonomous conversion of PE~to \mm~is characterized in Sec.~\ref{Sec:MachineOperation}

\section{Emergence of \mm} \label{Sec:EmWit_SO}

\begin{figure*}[t]
  \centering
  \includegraphics[width=0.99\linewidth]{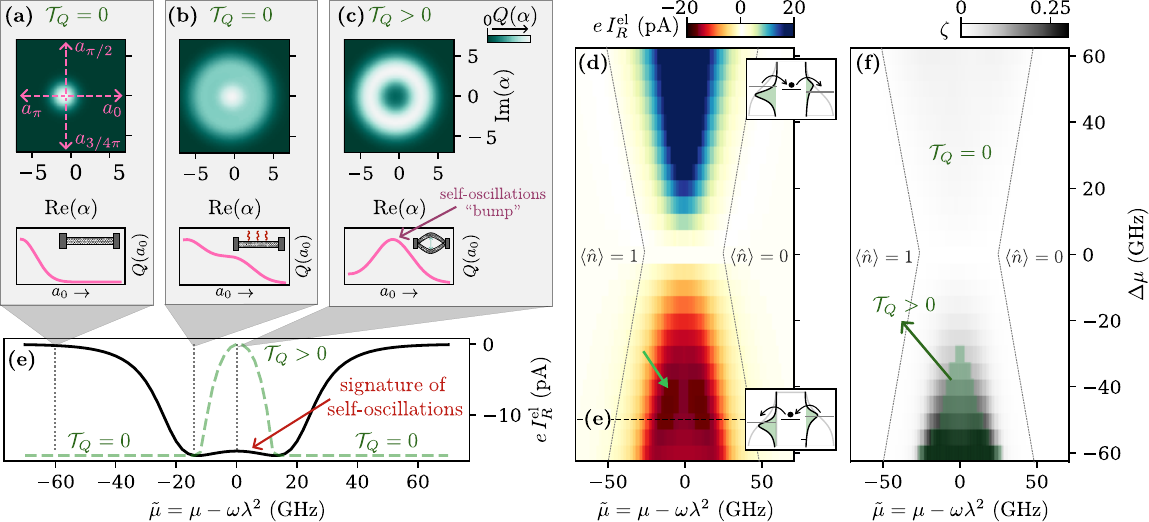}
  \caption{
  \textbf{Signatures and physical interpretation of the emergence of~\mm.}
  Panels \textbf{(a)}-\textbf{(c)} show the Husimi Q representation of \res's state (top) and its profile along the direction of $a_0$ (bottom). They correspond to the mechanical state of \textbf{(a)} no-motion, \textbf{(b)} heating, and \textbf{(c)} \mm.
  Panel \textbf{(d)} shows $\IR$ as a function of both the dot's normalized energy $\tilde{\mu}$ and chemical potential imbalance $\Delta\mu$. Blue values are for the left-to-right electronic hopping ($\IR>0$) and red values for the inverse direction ($\IR<0$). Insets indicate the direction of the particle current. 
  Gray lines delineate the Coulomb blockade half-diamonds: regions where there is no particle transport ($\IR=0$). The left half-diamond corresponds to the case where the \qd~is always occupied by an electron ($\langle \hat{n} \rangle =1$) while in the right half-diamond the \qd~is always empty ($\langle\hat{n} \rangle =0$). 
  For large values of $|\Delta \mu|$, a change of the monotonicity of the particle current is observed (highlighted with a green arrow). For clarity, in panel
  \textbf{(e)}, we plot a line cut of panel \textbf{(d)} at $\Delta\mu=-50$ GHz (black line).
  The region of change of monotonicity of the current coincides with the region where the \dnt~(dashed green) $\TT_Q>0$, indicating that the the resonator is self-oscillating. 
  Panel \textbf{(f)} plots the \mm~witness parameter $\zeta$ as a function of both $\tilde{\mu}$ and $\Delta\mu$. The green region is where \mm~are present based on a non-zero \dnt.
  Parameters: $\Gamma_L/(2\pi) = \Gamma_R/(2\pi) = 0.2$ GHz, $\omega/(2\pi) = 1$ GHz, $T_L=T_R=13.09$ GHz (which corresponds to $100$ mK), $\delta_L=\delta_R=10$ GHz, $\gamma_L=-\gamma_R=-10$ GHz, and $\lambda=0.7$.}
  \label{fig:Fig2}
\end{figure*}

\subsection{Characterizing \mm}
\label{sec:dntotropy}

Let us now study in detail the presence of periodic mechanical motion in the \res. This motion is referred to as \mm~\cite{Jenkins2013, Culhane2022, Wachtler2019, Vigneau2022, Wen2020, Tabanera2024}, with the prefix ``self'' emphasizing that the oscillatory motion is generated internally, due to the fixed \res--\qd~coupling and in the absence of any external time-dependent drive. Here, to maintain an intuitive connection to classical physics, we will work with phase-space distributions representing the state of the \res. Particularly, we use the experimentally obtainable Husimi q-representation~\cite{Schleich}
\begin{align}\label{eq:Husimi_function}
  \pi Q(\alpha) =\langle\alpha|\rho_\mathrm{\res}|\alpha\rangle, 
\end{align}
with $|\alpha\rangle$ being the coherent state with amplitude $\alpha$ and $\rho_\mathrm{\res}=\Tr_{\mathrm{\qd}}\left({\rho}_s\right)$ the \res~reduced density matrix in the \NESS~\footnote{We note that the use of the Husimi distribution is a choice, all the considerations in the text can be applied to any distribution function.}. In the top panels of Fig.~\subfigref{fig:Fig2}{a--c}, we show three representative cases of steady state distributions for this system. In Fig.~\subfigref{fig:Fig2}{a}, we observe a bell-shaped distribution, indicating that the most probable state of the resonator is the ground state, and there is no discernible periodic motion. Fig.~\subfigref{fig:Fig2}{b} is qualitatively similar with an enlargement of the distribution, which indicates that the resonator is ``hotter'' than in case Fig.~\subfigref{fig:Fig2}{a}. Finally, in panel Fig.~\subfigref{fig:Fig2}{c}, the distribution presents a ``torus'' (or ring) shape centered at $\alpha \sim 0$. Recalling that a classical oscillator draws a circle in the phase space, this form of distribution is associated with the presence of periodic motion and, following the literature~\cite{Gorelik_1998, Usmani_2007, Ludwig_2008, Wachtler2019, Urgell_2020, Wen2020, Culhane2022, Rodrigues2007}, the \res~is said to be self-oscillating. Naturally, the more defined and narrow the torus, the higher the quality of the oscillatory state.

To our knowledge, no faithful characterization of \mm, beyond visually checking whether the phase-space distribution is bell- or torus-shaped, has been developed. The closest attempt was made in Ref.~\cite{Culhane2022}, where ergotropy~\cite{Allahverdyan_2004} of the \res{} was proposed. However, in the quantum domain considered here, the ergotropy turns out to be an unfaithful indicator of \mm{} as bell-shaped distributions may also have nonzero ergotropy. This is due to the fact that ergotropy can arise for reasons unrelated to \mm, such as coherence build-up. We explore this numerically in Appendix~\ref{Appendix:doughnutropy}. 

In this paper, we introduce a novel rigorous measure of ``torus-shapedness'' of a phase-space distribution $Q$ of a \res{}, which we denote by $\TT_{Q}$ and call \dnt{}. Mathematically, we define it as
\begin{equation}\label{eq:doughnutropy}
  \TT_{Q} = \min_{\varphi \in [0, 2 \pi)} \frac{1}{S_{Q_\varphi}} \int_{a_\varphi} d\alpha \,|\alpha - \alpha_c| \big[Q_\varphi(\alpha) - Q_\varphi^\downarrow(\alpha)\big], 
\end{equation}
where $\alpha_c := \int d^2\alpha \, \alpha \, Q(\alpha)$ is the barycenter of the phase-space distribution and $a_\varphi$ is the infinite half-line originating at $\alpha = \alpha_c$ and oriented at angle $\varphi \in [0, 2\pi)$. The function $Q_\varphi \, : \, a_\varphi \to \mathbbm{R}_+$ is the normalized restriction of $Q(\alpha)$ to $a_\varphi$,
\begin{align}
    Q_\varphi(\alpha) := \frac{Q(\alpha)}{\int_{a_\varphi} d\bar{\alpha} \, Q(\bar{\alpha})} \bigg\vert_{\alpha \in a_\varphi} \, ,
\end{align}
and $Q_\varphi^\downarrow(\alpha)$ is its non-increasing rearrangement \cite{Lieb-Loss} along $a_\varphi$. Finally, $S_{Q_\varphi} := - \int_{a_\varphi} d\alpha \, Q_\varphi(\alpha) \ln (Q_\varphi(\alpha))$ is the entropy of $Q_\varphi$ \footnote{Not to be confused with the Wehrl entropy $\int d^2\alpha \, Q(\alpha) \ln Q(\alpha)$ \cite{wehrl_1979}.}. It is there to safeguard us from falsely inferring strong \mm{} in situations where $Q(\alpha)$ exhibits only weak non-monotonicities but at large values of $|\alpha - \alpha_c|$. 

Our definition of \dnt{} is inspired by the notion of ergotropy \cite{Allahverdyan_2004}, which is the rationale behind the name. Ergotropy quantifies by how much a system's average energy can be reduced by manipulating its state unitarily. Similarly, given a torus-like shape, \dnt{} quantifies by how much the shape's average radius can be reduced by measure-preserving operations (which are analogous to unitary operations). More on the structural similarity of the two can be found in Appendix~\ref{app:ergo_vs_toro}.

Defined as above, the \dnt{} faithfully captures the onset of \mm{} of the \res. Indeed, $\TT_{Q} > 0$ if and only if the distribution is not bell-shaped in that it presents a ``bump'' along all half-lines $a_\varphi$ (mind the minimization in Eq.~\eqref{eq:doughnutropy}). A visual demonstration of this is provided in Fig.~\subfigref{fig:Fig2}{a--c}, where the bottom panels show the vertical cross sections of the distributions above them. Throughout the work, the center of the distribution is placed at $-\lambda\av{\hat{n}}$, which is motivated by the fact that $\alpha_c \approx -\lambda\langle\hat{n}\rangle$ across the full range of coupling strengths considered here. We demonstrate this in Appendix~\ref{Appendix:doughnutropy}, which also contains an extended discussion on $\TT_{Q}$. 

In our simulations, we found that, for $\lambda\leq 1$, all distributions are rather symmetric with respect to their barycenters. Therefore, for numerical tractability, we carry out the minimization in Eq.~\eqref{eq:doughnutropy} only over the half-lines $\{ a_0, a_\pi, a_{\pi/2}, a_{3\pi/4}\}$ shown on the top panel of Fig.~\subfigref{fig:Fig2}{a}. For $\lambda \gg 1$, a larger set of $\varphi$'s must be used to achieve a reliable minimization. Apart from that, the situation for $\lambda \gg 1$ is qualitatively the same as for $\lambda \lesssim 1$.

\subsection{Witnessing \mm}

Although \dnt{} is a precise measure of \mm{}, it is obviously hard to access experimentally as it requires the full knowledge of the state. Here we propose two experiment-friendly approaches to witnessing \mm{}. To do so, let us first present an analogy of our device with an everyday example of the \mm~phenomenon: water flowing through a garden hose. A slow flow of water does not move the hose. However, as the flow increases, after some threshold, the garden hose starts to move and falls into an oscillatory pattern. This resembles the platform studied here: the electronic flow takes the place of the water flow and the resonator takes the place of the garden hose.

As this analogy vividly illustrates, the electronic particle current is a key parameter controlling the emergence of \mm; if there is no particle current through the \qd~then no motion is possible in the \res. To evidence this, we consider the current flowing from reservoir $\nu$ given by~\cite{Schaller2014Book, Leijnse2010, Potts2024} 
\begin{align} \label{def:particle_current}
  \I &= \partial_t \langle \hat{N_\nu} \rangle,
\end{align}
where $\langle \hat{N}_\nu \rangle= \Tr(\c^\dagger \c {\rho})$ is the average particle number of the reservoir $\nu$. The sign convention is such that currents are positive if the particle flows towards the reservoir $\nu$, and negative otherwise~\cite{Potts2024,Humphrey2002}. An explicit expression for $\I$ for the present model is derived in Ref.~\cite{Sevitz2025}. 

Here we focus on the \NESS~where the total particle current is conserved, $\IL + \IR = 0$. Therefore, it is sufficient to consider only $\IR$. For simplicity, let us for now consider the case of $\Delta T = T_L - T_R = 0$ mK, where transport is solely due to chemical imbalance $\Delta \mu$, as $\mu_L=+\Delta\mu/2$ and $\mu_R=-\Delta\mu/2$. Here and throughout, we choose the experimentally feasible values of $\omega /(2\pi) = 1$ GHz for the frequency of the \res{} and $\Gamma_L /(2\pi) = \Gamma_R /(2\pi) = 0.2$ GHz for the tunneling rates~\cite{Laird2012, Samanta2023}. These rates are an order of magnitude lower than $\omega$, and therefore the commonly-employed semi-classical approach is no longer valid, and thus the fully quantum modeling presented in Ref.~\cite{Sevitz2025} is required.

In Fig.~\subfigref{fig:Fig2}{b}, we plot the numerically obtained current $\IR$ as a function of both $\tilde{\mu}$ and $\Delta\mu$ for a fixed coupling $\lambda=0.7$. There, between the Coulomb blockade half-diamonds, electronic transport is enabled, with hopping from left to right for $\Delta\mu > 0$ and in the inverse direction for $\Delta\mu < 0$. In the negative bias region, we see a sharp decrease of the current (see the green arrow). In Fig.~\subfigref{fig:Fig2}{e}, we plot a line cut at $\Delta\mu= -50 $ GHz for clarity. This change of monotonicity behavior is a key experimental witness of \mm~commonly referred to as the switching effect~\cite{Eichler2011, Schmid2015, Schmid2012, Willick2020, Tabanera2024}, which has been extensively observed experimentally in the classical regime~\cite{Wen2020,Eichler2011, Schmid2015, Schmid2012, Willick2020, Tabanera2024}. To see whether this is also the case in the quantum domain, in dashed green, we plot the \dnt{}. Note that the only region where $\TT_Q>0$ corresponds to the region where the switching effect appears in the current. With this, we can conclude that a change of monotonicity in the current can be used as a witness of \mm~in the \res~with a high degree of confidence. To our knowledge, this is the first time a model exhibits that the switching effect is related to \mm~in the quantum slow transport regime for this device with arbitrary \qd--\res~coupling, going beyond Ref.~\cite{Rodrigues2007}.

\medskip

An alternative, more quantitative, approach also leverages the average energy of the \res{}, $E^\mathrm{mec} = \omega \av{\hat{N}_{\mathrm{ph}}}$, with $\av{\hat{N}_\mathrm{ph}} = \Tr(\bb\b \, \rho_\mathrm{\res})$, along with $|\I|$. Based on the simple fact that \mm{} cannot be generated without increasing the energy of the \res{}, we found that the adimensional parameter
\begin{align}\label{eq:SO_condition_R}
    \zeta := \frac{|\IR| E^\mathrm{mec}}{\omega^2} = \frac{|\IR| \av{\hat{N}_{\mathrm{ph}}}}{\omega}
\end{align}
is a good predictor of \mm{}. This is demonstrated in Fig.~\subfigref{fig:Fig2}{f}, where we plot $\zeta$ as a function of both $\tilde{\mu}$ and $\Delta\mu$, again for coupling $\lambda=0.7$. There, the region where \mm~occur, as dictated by $\TT_Q > 0$, is shown in green, and this region corresponds to larger values of $\zeta$.

\section{Powering \mm~by electronic transport}\label{Sec:MachineOperation}

\begin{figure*}[t]
  \centering
  \includegraphics[width=1\linewidth]{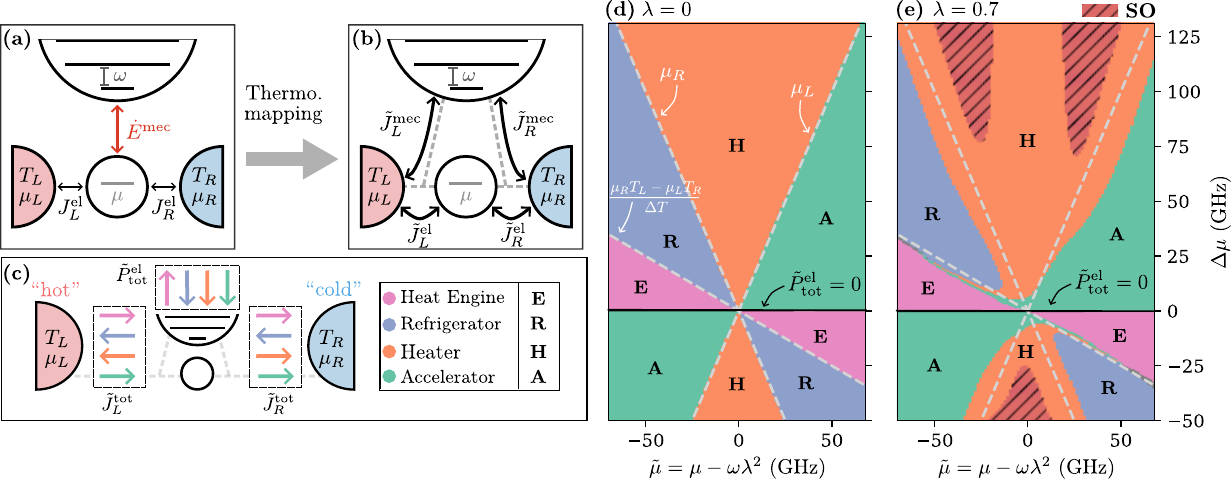}
  \caption{\textbf{Particle exchange to \mm~operation.}
  \textbf{(a)} Sketch of \qd+\res~platform in the original frame. Two types of heat currents can be identified: one between the \qd~and reservoir $\nu= {L,R}$, label as $J^\mathrm{el}_\nu$, and another between the \qd~and the \res, labeled as $\dot{E}^\mathrm{mec}$.
  \textbf{(b)} Platform after the application of the thermodynamic mapping. Here the \qd--lead coupling and separately the \qd--\res~coupling are transformed into a single three-body coupling (\qd+\res+lead $\nu$) leaving evident that whenever an electron hops from the reservoir $\nu$ to the \qd. Here energy is exchanged with both the \qd~as $\J^\mathrm{el}$ and also the \res~as $\J^\mathrm{mec}$. 
  \textbf{(c)} Schematic of relevant operating modes. The color coding is the following; pink Heat Engine (\textbf{E}), blue Refrigeration (\textbf{R}), orange Heater (\textbf{H}) and with green Accelerator (\textbf{A}). The dashed gray lines mark the analytical operation bounds of a PE machine. 
  Operation modes of the PE autonomous machine for coupling $\lambda=0$ \textbf{(d)} and $\lambda=0.7$ \textbf{(e)}. With gray dotted lines we denote the analytical bounds of operation. In panel \textbf{(e)} the region where \mm~(\textbf{SO}) are present are highlighted in red with diagonal black lines. 
  Parameters: $\Gamma_L/(2\pi) = \Gamma_R/(2\pi) = 0.2$ GHz, $\omega/(2\pi) = 1$ GHz, $T_L = 13.09$ GHz ($100$ mK), $T_R = 7.86$ GHz ($60$ mK) and $\delta_L=\delta_R=10$ GHz and $\gamma_L=-\gamma_R=-10$ GHz.
  }
  \label{fig:Fig3}
\end{figure*}

In the previous section, we evidenced that the electronic particle current enables an energy transfer to (or from) the \res~that may lead to \mm. Here we quantify energetic currents that modify the system's energy; see Fig.~\subfigref{fig:Fig3}{a}. Before the steady state is reached, there is a nonzero energy injection $\dot{E}^\mathrm{mec}$ from the \qd~to \res~and vice versa. But once the \NESS~is achieved, the \qd$\to$\res{} and \res$\to$\qd{} energy currents balance out, resulting in $\dot{E}^\mathrm{mec}=0$. However, knowing that the energy flow between the \qd~and lead $\nu$ remains nonzero, even in the \NESS, we expect the same for the energy flow that travels from the lead $\nu$ to the \res~(mediated by the \qd). Therefore, for a more transparent bookkeeping, we will switch to a picture where the energy current flowing through the \res~comes directly from the leads, which we expect to not weaken over time. In the following, we apply a thermodynamic mapping which achieves what we are seeking.

\subsection{Thermodynamic mapping}

The mapping is achieved by applying the polaron transformation~\cite{Landau1948, Holstein1959_I, Holstein1959_II, Cheng2008}, defined as $\tilde{\bullet}=P \, \hat{\bullet} \, P^\dagger$, to Eq.~\eqref{eq:H_total}. Here, $P=e^{\lambda (\hat{b}^\dagger-\hat{b}) \hat{d}^\dagger\hat{d}}$ and $\hat{\bullet}$ designates an arbitrary operator~\cite{Siddiqui2007,Mahan2000}. Under this transformation, the reservoir coupling introduced in Eq.~\eqref{eq:V_r} becomes~\cite{Siddiqui2007, Schaller2014Book}
\begin{equation}\label{eq:V_r_tilde}
  \tilde{V} = \sum_{\nu} t_\nu \c^\dagger \d e^{-\lambda (\b^\dagger-\b)} + \mathrm{h.c.}.
\end{equation}
This three-body (\qd~+ \res~+ reservoir $\nu$) form of the coupling makes it evident that every single electron transport from reservoir $\nu$ to the \qd{} induces momentum transfer to the \res. 

Furthermore, in this picture, the system Hamiltonian introduced in Eq.~\eqref{eq:H_s} is reduced to~\cite{Siddiqui2007,Piovano2011,Schaller2014Book}
\begin{equation}\label{eq:H_s_tilde}
  \tilde{H}_s=\tilde{\mu} \d^\dagger \d + \omega \bb \b, 
\end{equation}
where the \qd--\res~coupling modulated by $\lambda$ is absorbed as a shift in \qd's chemical potential $\tilde{\mu}=\mu-\omega \lambda^2$. This absorption of the \qd--\res~coupling enables us to separate the energetic contributions of the electronics from the mechanics. In Fig.~\subfigref{fig:Fig3}{b}, we provide a visual schematic of the effect of the this mapping.

\subsection{Thermodynamics}

First, let us identify all the energetic contributions that modify the system's energy. Eq.~\eqref{eq:H_s_tilde} suggests that, in the polaron frame, the electronic and mechanic energies separate. This is an advantage of the fully quantum-mechanical treatment of the resonator: this contribution separation was not possible when the oscillator was considered to be classical~\cite{Wachtler2019}. Making use of the first law of thermodynamics (see Appendix~\ref{Appendix:Engine_thermo}), we obtain
\begin{equation}\label{eq:first_law_thermo}
  \partial_t \langle {H}_s \rangle = \partial_t \langle \tilde{H}_s \rangle_P =  - \sum_\nu \big( \J^\mathrm{el} + \J^\mathrm{mec} \big) - \tilde{P}^\mathrm{el}_\mathrm{tot}, 
\end{equation}
with $\langle \bullet \rangle= \Tr\left( \bullet \rho_s\right)$ and $\langle \bullet \rangle_P= \Tr\left( \bullet \trho_s\right)$ the steady-state expectation values in, respectively, the original and polaron frames. Here, we identify two heat-like currents: one associated with the electrical current (``$\mathrm{el}$'') and another one related to the energy exchange with the \res~due to electron hopping (``$\mathrm{mec}$''). The tilde is to clarify that we are defining them in the polaron frame. These energy currents are depicted in Fig.~\subfigref{fig:Fig3}{b}.

On the one hand, the electrical heat current, defined as 
\begin{equation}\label{eq:J_el}
  \J^\mathrm{el}= (\mu-\mu_\nu) \I ,
\end{equation}
with the particle current $\I$ defined in Eq.~\eqref{def:particle_current}, measures the energy the electron transports from reservoir $\nu$ to the \qd. Note that it has the ``typical'' form in the sense that, if we set the \qd--\res~coupling to zero ($\lambda=0$), this expression reduces to the widely studied \qd~electrical heat currents~\cite{Lu2023, Manzano2020, Schaller2014Book, Leijnse2010, Sanchez2011}. On the other hand, we find that the mechanical heat current in Eq.~\eqref{eq:first_law_thermo} is
\begin{align}\nonumber
  \J^\mathrm{mec} &= - \sum_{jkl}  \omega j \Big( \Qnn{0}{1}{j}{j}{k}{l} \bra{k} \trho_s^{0}\ket{l} -  \Qnn{1}{0}{j}{j}{k}{l} \bra{k} \trho_s^{1}\ket{l} \Big) \\ \label{eq:J_mec} & \quad \quad \quad \quad \quad \quad - \lambda^2 \omega \I .
\end{align}
The first term represents the total phononic energy flow modulated by the tensors 
\begin{align}\label{eq:Q_01}
  \Qnn{0}{1}{j}{m}{k}{l}  
  &\propto R^{0\rightarrow 1}_\nu = \G(\epsilon) f_\nu(\epsilon)
  \\ \label{eq:Q_10}
  \Qnn{1}{0}{j}{m}{k}{l}  
  &\propto R^{1\rightarrow 0}_\nu = \G(\epsilon) [1 - f_\nu(\epsilon)], 
\end{align}
with $\G$ the energy-dependent tunneling rate of lead $\nu$ and $f_\nu$ its Fermi function. See Appendix~\ref{Appendix:Engine_thermo} for the full expressions. The second term of Eq.~\eqref{eq:J_mec} plays the same role as the conserved quantity $\I$ in $\J^\mathrm{el}$~\cite{Manzano2020}, but here instead of being weighted by the chemical potential of the reservoir $\nu$, it is weighted by $\lambda^2 \omega$ related to the \qd--\res~coupling. In total, the quantity measures the heat flow to (from) the reservoir $\nu$ due to the electronic tunneling. By construction, it is strictly zero for the un-coupled case $\lambda=0$, see Appendix~\ref{Appendix:HeatCurrents} for further details. It is worth emphasizing that, unlike previous works~\cite{Wachtler2019}, we are able to access this heat current strictly because we are in the fully quantum domain and the polaron frame allows us to separate the heat current contribution.

Finally, the last component of Eq.~\eqref{eq:first_law_thermo} is the total power~\cite{Manzano2020}
\begin{equation}\label{eq:total_el_power}
    \Power = -\Delta\mu \,\IR.
\end{equation}
Since no external drive is acting on the device, the only power that can exist in this platform is electrical, and it is provided by the chemical imbalance~\cite{Potts2024}. Thus, Eq.~\eqref{eq:total_el_power} quantifies the power generated from the electronic hopping that drives a particle current against (or in favor of) the lead's chemical bias~\cite{Mayrhofer2021,Sothmann2015,De2016}. 

\subsection{Machine operation}

With this we have all the necessary ingredients to study the operation of the 
autonomous machine that converts PE into \mm~in the attached \res. We take the same experimental feasible parameters as in the previous section, but now we also consider the cases where the temperature bias is not zero, specifically $T_L>T_R$. This is to identify the different regimes of operation of the thermodynamics machine. Making use of the total heat current from the right (left) $\JR^\mathrm{tot}=\JR^\mathrm{el}+\JR^\mathrm{mec}$ ($\JL^\mathrm{tot}=\JL^\mathrm{el}+\JL^\mathrm{mec}$) reservoir, we identify four relevant operating modes highlighted in Fig.~\subfigref{fig:Fig3}{c}: heat engine (energy is transferred from the hot reservoir to the cold reservoir while generating an electrical power output), refrigerator (power is used to extract heat from the cold reservoir to the hot reservoir), heater (power is used to provide heat to both the cold and the hot reservoir) and accelerator (power is used to help heat transfer from the hot to the cold reservoir). 

For reference, we first look at the un-coupled \qd--\res~case ($\lambda=0$) which represents the typical PE machine schemes considered before~\cite{Humphrey2002, Mahan1996, Esposito2009, Nakpathomkun2010, Borga2012, Esposito2012,Josefsson2018}. In Fig.~\subfigref{fig:Fig3}{d} we plot the phase diagram of the different machine operations as a function of the dot's re-normalized energy $\tilde{\mu}$ and the reservoirs chemical imbalance $\Delta \mu$. We delimit each operating mode and with dashed gray lines we denote the analytical boundaries given in Ref.~\cite{Potts2024}. Now, in Fig.~\subfigref{fig:Fig3}{e}, we showcase the same phase diagram as in Fig.~\subfigref{fig:Fig3}{d} but for a \qd--\res~coupling of $\lambda=0.7$. It is interesting to note how the machine operation boundaries are modified due to the coupling; the heater regime seeps onto the neighboring regions of operation~\footnote{We observe some regions close to the boundaries (marked with gray) that do not correspond to any operation mode, we attribute this to a numerical artifact.}. On top of this plot we have delimitated the region where \mm~occurs (denoted by \textbf{SO}) using the criteria detailed in the previous section, see Eq.~\eqref{eq:doughnutropy}. Here we observe \mm~for positive and negative values of $\Delta\mu$, see Appendix~\ref{Appendix:TransportProp_Delta_T_40} for the transport quantities. The regime where our steady state autonomous machine converts particle exchange into \mm~falls with the machine operating as a heater. That is, the power provided by the chemical bias $\Delta \mu$ is used to heat both the hot and cold reservoir. From a fundamental perspective, this observation is a motivation to explore other operation regimes apart from the commonly studied heat engines and refrigerators. Here we give an application to the heater regime, which is generally discarded as ``un-useful''. In the next, and final, section we analyze the performance of the \mm~generation.

\subsection{Performance metric}
\label{Sec:Performance}

\begin{figure}[t]
  \centering
  \includegraphics[width=\linewidth]{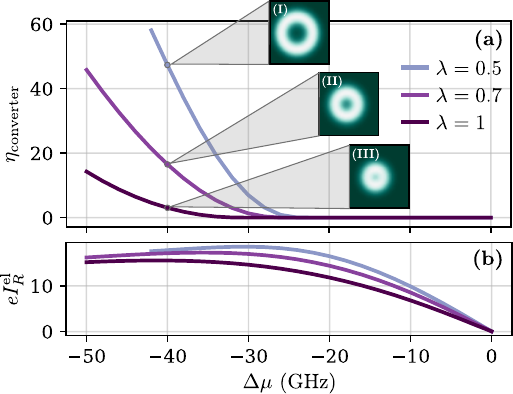}
  \caption{\textbf{Performance of the particle-exchange to \mm~autonomous quantum machine.} 
  (\textbf{a)} Machine performance $\eta_\mathrm{converter}$ as a function of the chemical imbalance $\Delta\mu$ with a fixed re-normalized energy $\tilde{\mu}=0$ GHz for \qd--\res~couplings $\lambda=0.5$, $\lambda=0.7$, and $\lambda=1$.
  In the insets (I)-(III) we compute the Husimi distribution $\Delta \mu= -40$ GHz with couplings (I) $\lambda=1$, (II) $\lambda=0.7$, and (III) $\lambda=1$.
  \textbf{(b)} $\IR$ as function of the $\Delta\mu$ for the same $\lambda$ couplings as panel \textbf{(a)}. 
  Parameters: $\Gamma_L/(2\pi)=\Gamma_R/(2\pi)=0.2$ GHz, $\omega/(2\pi)=1$ GHz, $T_L= 13.09$ GHz ($100$ mK), $T_R=7.86$ GHz ($60$ mK) and $\delta_L=\delta_R=10$ GHz, $\delta_L=\delta_R=10$ GHz, $\gamma_L=-\gamma_R=-10$ GHz and $\tilde{\mu}=0$ GHz.}
  \label{fig:Fig4}
\end{figure}

In this last section we quantify how \textit{well} the autonomous quantum machine converts PE to \mm. For this, we define the experimentally accessible performance measure
\begin{equation}\label{eq:Performance_converter}
  \eta_\mathrm{converter}= \frac{\omega \TT_Q}{|\IR|} \geq 0.
\end{equation}
Here we used that in the \NESS~$|\IR|=|\IL|$ and include $\omega$ to make the measure dimensionless. By construction, it will only yield nonzero values if the resonator is self-oscillating ($\TT_Q > 0$). In Fig.~\subfigref{fig:Fig4}{a}, we compare the performance measure as a function of the chemical imbalance $\Delta \mu$ for different \qd--\res~couplings and a fixed re-normalized dot energy of $\tilde{\mu}=0$ GHz. For a given $\Delta \mu$, the lower the coupling the higher the performance. For reference, the insets {(I)}-{(III)} showcase the Husimi distribution for different couplings at $\Delta \mu = -40$ GHz. We attribute this feature to the Franck--Condon current blockade where the larger the coupling $\lambda$ is, the more $|\IR|$ is reduced, see Fig.~\subfigref{fig:Fig4}{b}~\cite{Koch2005,Leijnse2010,Leturcq2009,Koch2006}. This reduction of the particle current in turn induces less energy to the \res, thus reducing the quality of the \mm.

Lastly, we remind the reader that in our setup the \res~is not coupled to an external bath. However, in a realistic experimental scenario, a tradeoff between the external damping and the \qd--\res~coupling will define an optimal situation. This will be the topic of future work.

\section{Discussions and conclusion}\label{Sec:Conclusion}

We have theoretically studied the operation of an autonomous quantum machine that converts PE hosted in a \qd{} into \mm{} by coupling the dot to a \res{} work load. In contrast to previous studies, we treated the system in the never-before-looked-at slow tunneling regime, where a semiclassical approach is no longer justified and thus a fully quantum model is required. 

First, we have shown that, in this regime, \mm{} of the \res{} arise due to energy provided via the particle exchange mechanism for arbitrary \qd--\res{} coupling. Furthermore, we defined an ergotropy-inspired measure that faithfully distinguishes and quantifies \mm. Using this rigorous tool, we demonstrated that \mm{} can be witnessed by the electrical particle current, thereby recovering and extending the well-known switching effect to this regime.  Additionally, we provided two alternative witnesses of the emergence of \mm.

Next, we analyzed the thermodynamics of the machine in the \NESS. We found that the polaron transformation provides a thermodynamic mapping that separates the electrical and mechanical ``channels'' of the total energy current through the system. We discovered that, under realistic conditions, \mm{} only occur when the machine operates as a heater, i.e., when the electrical power is used to heat both the hot and cold fermionic reservoirs. 

Finally, to characterize the performance of the machine, we defined an efficiency of the conversion of electrical current into \mm. Doing so revealed that lower \qd--\res{} couplings yield better quality of \mm. We note that the last finding likely holds only when the \res{} is uncoupled from the environment. Realistically, we expect a tradeoff among the external damping acting on the \res{} and the \qd--\res{} coupling. Determining the optimal performance of our setup in this scenario is an interesting direction for future research.

Balancing experimental feasibility and theoretical simplicity, our setup provides a gateway into heat and work exchanges in realistic nanoscale quantum devices. With this study we hope to encourage the community to study work extraction and storage using this and similar platforms that couple to quantum-mechanical work loads. As a future direction, we are looking to exploit other devices such as NV centers coupled to cantilevers~\cite{Arcizet2011,Oeckinghaus2020} and spin--mechanical coupling in carbon nanotubes~\cite{Fedele2024}.


\begin{acknowledgments}
The authors would like to acknowledges useful conversations with Andrew N. Jordan, Jorge Tabanera-Bravo, Gernot Schaller, Christopher W. Wächtler, Juliette Monsel, Federico Fedele, Milton Aguilar, and Felix Hartmann. S.S. and J.A. acknowledge the support by Deutsche Forschungsgemeinschaft (DFG 384846402). J.A. and F.C. acknowledge support from EPSRC (EP/R045577/1). J.A. thanks the Royal Society for support. J.A. and K.V.H. are grateful for support from the University of Potsdam. Computation of the non-equilibrium steady states were performed with computing resources provided by ZIM, Universität Potsdam.
\end{acknowledgments}

\bibliography{bibliography.bib}

\appendix

\onecolumngrid

\section{Master equation}
\label{Appendix:Master_equation}

In this section we summarize the main expressions of Ref.~\cite{Sevitz2025} used in this work. The application of the derivation done in that work is valid under the following assumptions:
\begin{itemize}
    \item[(i)] Single electron tunneling regime where $\Gamma_\nu \ll T_\nu$.
    \item[(ii)] Initially ($t_0=-\infty$) the system and the reservoirs are un-correlated $\trho_{s}(t_0)\otimes \rho_\R$, where $\rho_\R =  \bigotimes_{\nu} \frac{1}{Z_\nu}\exp\{- \beta_\nu\left((\epsilon_{\nu}-\mu_\nu) \c^\dagger \c\right)\}$ is the density matrix of the reservoirs at thermal equilibrium and $\trho_{s}(t_0)$ is the initial state of the \qd+\res~system.
    \item[(iii)] The leads are infinite such that they are not affected by the electronic tunnelings and are always in thermal equilibrium. 
    \item[(iv)] The Markov approximation is valid, where the system dynamics at time $t$ is determined \textit{only} by the operators at the same time. This is always valid in the \NESS~considered here as long as the bath correlation function decay at some time scale $t<+\infty$, we return to this point the following subsection.
    \item[(v)] As the tunneling rates maintain their energy dependence the additional condition $\delta_\nu\gg \Gamma_\nu$ is needed.
\end{itemize}
Under the approximations (i)-(v) the \NESS~is obtained by the null space of the following master equation 
\begin{equation}\label{Eq_app:ME_Redfield_final}
  \bra{j}\partial_t\trho_s^n \ket{m} = -i\omega(j-m)\bra{j} \trho_s^n \ket{m} + \sum_\nu \sum_{kl}
  \Big( -\Rnn{n,}{n}{j}{m}{k}{l} \bra{k} \trho_s^n \ket{l}
  + \Rnn{n+1,}{n}{j}{m}{k}{l} \bra{k} \trho_s^{n+1} \ket{l}
   + \Rnn{n-1,}{n}{j}{m}{k}{l} \bra{k} \trho_s^{n-1} \ket{l} \Big).
\end{equation}
The shorthand notation $ \bra{k, n} \trho_s \ket{l, n} =\bra{k} \trho_s^n \ket{l}$ was used with the dot occupations $n=0,1$. Also, in Eq.~\eqref{Eq_app:ME_Redfield_final}, the Redfield tensors corresponding to lead $\nu$ were introduced as
\begin{align} \label{Eq_app:A00}
  \Rnn{0}{0}{j}{m}{k}{l} & = \frac{1}{2} \sum_i \Big( R^{0\rightarrow 1}_\nu(\e{k}{i}) \DD_{j,i}\D_{i,k} \delta_{ml} + R^{0\rightarrow 1}_\nu(\e{l}{i}) \DD_{l,i}\D_{i,m} \delta_{kj} \Big)
  \\ \label{Eq_app:A01}
  \Rnn{0}{1}{j}{m}{k}{l} &= \frac{1}{2}\D_{j,k}\DD_{l,m} \Big( R^{0\rightarrow 1}_\nu(\e{k}{j}) + R^{0\rightarrow 1}_\nu(\e{l}{m})\Big)
  \\ \label{Eq_app:A11}
  \Rnn{1}{1}{j}{m}{k}{l} & = \frac{1}{2} \sum_i 
  \Big( R^{1\rightarrow 0}_\nu(\e{i}{k}) \D_{j,i} \DD_{i,k}\delta_{ml} + R^{1\rightarrow 0}_\nu(\e{i}{l}) \D_{l,i} \DD_{i,m} \delta_{jk} \Big)
  \\ \label{Eq_app:A10}
  \Rnn{1}{0}{j}{m}{k}{l} &= \frac{1}{2} \DD_{j,k}\D_{l,m} \Big(R^{1\rightarrow 0}_\nu(\e{j}{k}) + R^{1\rightarrow 0}_\nu(\e{m}{l}) \Big)
\end{align}
where $\e{k}{l}=\tilde{\mu}-\omega(k-l)$, $\D_{kl} = \bra{k}e^{\lambda (\bb-\b)} \ket{l}$ and $R^{0\rightarrow 1}_\nu/R^{1\rightarrow 0}_\nu$ are the electronic transition rates defined as
\begin{align} \label{Eq_app:QD_R01}
  R^{0\rightarrow 1}_\nu(\epsilon) &= \G(\epsilon) f_\nu(\epsilon)
  \\ \label{Eq_app:QD_R10}
  R^{1\rightarrow 0}_\nu(\epsilon) &= \G(\epsilon) [1 - f_\nu(\epsilon)].
\end{align}
Recall that $f_\nu(\epsilon) = 1/[\exp(\beta_\nu (\epsilon - \mu_\nu)) + 1]$ is the Fermi distribution of reservoir $\nu$ with chemical potential $\mu_\nu$ and inverse temperature $\beta_\nu=1/T_\nu$ and $\G$ is the tunneling rates with energy dependence taken to be a Lorentzian profile, see Sec.~\ref{Sec:PlatformDescription} of the main text.

\subsection{Convergence of bath-correlation function}
\label{Appendix:Convergance_C}

\begin{figure}[htbp]
  \centering
  \includegraphics[width=.8\linewidth]{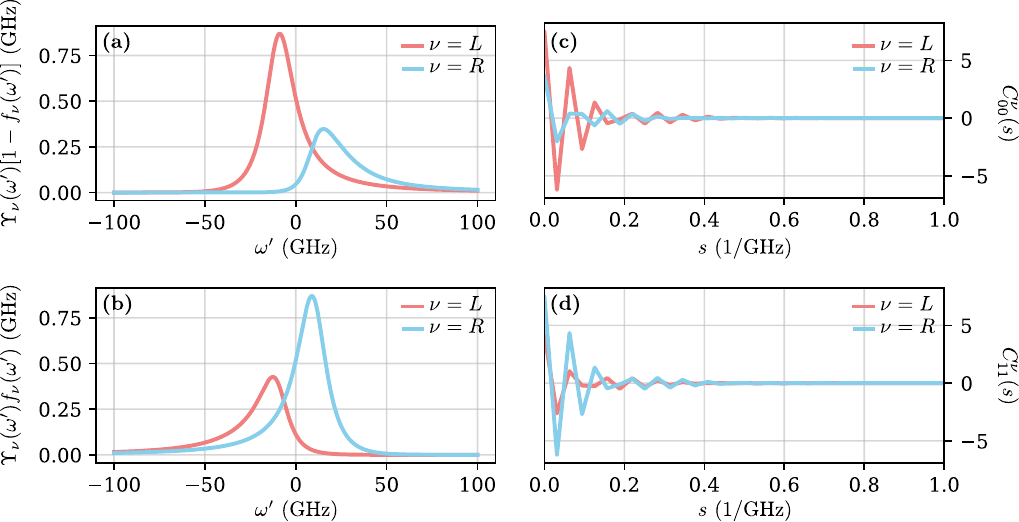}
  \caption{
  \textbf{(a)}-\textbf{(b)} Integration functions of the bath correlation functions as a function of frequency $\omega'$ 
  \textbf{(c)}-\textbf{(d)} Bath correlation functions detailed in Eq.~\eqref{Eq_app:C00} and \eqref{Eq_app:C11} respectively as a function of time $s$. 
  Parameters: $\Gamma_L/(2\pi)=\Gamma_R/(2\pi)= 0.2$ GHz, $\omega/(2\pi)=1$ GHz, $T_L= 13.09$ GHz ($100$ mK), $T_R=7.86$ GHz ($60$ mK) and $\delta_L=\delta_R=10$ GHz, $\mu_R=-\mu_L=20$ GHz, and $\gamma_L=-\gamma_R=-10$ GHz.}
  \label{fig:Fig5_appendix}
\end{figure}

The master equation used to compute the \NESS~of the system detailed in Appendix.~\ref{Appendix:Master_equation} is local in time, that is the state at time $t$ is only determined by the state at the same time~\cite{Sevitz2025}. This is justified is the bath correlation time $\tau_\R$ is smaller than any relevant time-scale of the system~\cite{Potts2024}. Though as we are only interested in the \textit{long time scales} (the \NESS) of the system, to assume time-local it is sufficient to check that bath correlation functions effectively decay at some time~\cite{Strasberg2022Book}. Here, we expose that for the parameters chosen in this work this condition is satisfied. For this we first write down the bath correlation functions
\begin{align} \label{Eq_app:C00}
   C^\nu_{00}(s) &= \frac{1}{2\pi}\int_{-\infty}^\infty d\omega' \, e^{-i \omega' s} \G(\omega') [1 - f_\nu(\omega')]
   \\ \label{Eq_app:C11}
   C^\nu_{11}(s) & =  \frac{1}{2\pi} \int_{-\infty}^\infty d\omega' \, e^{+i\omega' s} \G(\omega') f_\nu(\omega'),
\end{align} 
see Ref.~\cite{Sevitz2025}. In Fig.~\ref{fig:Fig5_appendix} we showcase an example case that represents the behavior of the parameters used in this work. In Fig.~\subfigref{fig:Fig5_appendix}{a--b} we plot the integration function of Eq.~\eqref{Eq_app:C00} and \eqref{Eq_app:C11} respectively. We integrate these functions numerically using the Qutip "spectrum\_correlation\_fft" function~\cite{Lambert2024Qutip5} to obtain the correlation function detailed in Eq.~\eqref{Eq_app:C00} and \eqref{Eq_app:C11}. These are plotted in Fig.~\subfigref{fig:Fig5_appendix}{c--d} respectively. Note that after some oscillations, all the function decay to zero. Thus this justifies the use of the Redfield equation used to compute the \NESS~of the system for the parameters used in the paper.

\section{More on the geometry of the Husimi distribution and \dnt{}}
\label{Appendix:doughnutropy}

\subsection{Barycenter of the Husimi function}
\label{app:barycenter}

Let us rewrite the barycenter of $Q(\alpha)$ as
\begin{align}
    \alpha_c = \int d^2 \alpha \, \alpha Q(\alpha) = \Tr\Big(\rho_{\mathrm{\res}} \int \! \frac{d^2\alpha}{\pi} \alpha \ket{\alpha}\!\bra{\alpha} \Big) = \Tr\Big(\rho_{\mathrm{\res}} \, \hat{b} \int \! \frac{d^2\alpha}{\pi} \ket{\alpha}\!\bra{\alpha} \Big) = \Tr\big(\rho_{\mathrm{\res}} \, \hat{b}\big).
\end{align}
Keeping in mind that $\rho_{\mathrm{\res}} = \Tr_{\mathrm{\qd}}(\rho_s)$, we have that $\Tr(\rho_{\mathrm{\res}} \, \hat{b}) = \Tr(\rho_s \, \hat{b})$ where, on the right-hand side, the trace is over the full \res+\qd{} Hilbert space and by $\hat{b}$ we mean $\hat{b} \otimes \id_{\mathrm{\qd}}$, with $\id_{\mathrm{\qd}}$ the identity operator in \qd's Hilbert space (note that we suppress tensor notation throughout this paper). Thus,
\begin{align}
    \alpha_c = \Tr(\rho_s \hat{b}).
\end{align}

To gather further insight into $\alpha_c$, let us transition to the polaron frame, where $\alpha_c = \Tr(\tilde{\rho}_s \, P \hat{b} P^\dagger)$. Keeping in mind that $P = e^{\lambda \hat{b}^\dagger - \lambda^* \hat{b}}$ and using the Campbell identity, we readily find that $P \hat{b} P^\dagger = \hat{b} - \lambda \hat{n}$, leaving us with
\begin{align} \label{Eq_app:esor1}
    \alpha_c = \Tr(\tilde{\rho}_{\mathrm{\res}} \, \hat{b}) - \lambda \Tr(\tilde{\rho}_s \, \hat{n}) \stackrel{(*)}{=} \Tr(\tilde{\rho}_{\mathrm{\res}} \, \hat{b}) - \lambda \Tr(\rho_s \, \hat{n}) = \Tr(\tilde{\rho}_{\mathrm{\res}} \, \hat{b}) - \lambda \av{\hat{n}},
\end{align}
where step ($\ast$) is due to the fact that $[P, \hat{n}] = 0$. Recalling that, in the Fock basis $\{\ket{k}\}_{k=0}^\infty$ of the \res, $\hat{b}\ket{k} = \sqrt{k}\ket{k-1}$, we can rewrite Eq.~\eqref{Eq_app:esor1} as
\begin{align}
    \alpha_c = \sum_{k = 1}^\infty \sqrt{k} \left(\tilde{\rho}_{\mathrm{\res}}\right)_{k, k-1} - \lambda \av{\hat{n}}.
\end{align}
Now, looking at Eqs.~\eqref{eq:V_r_tilde} and~\eqref{eq:H_s_tilde}, we see that $\tilde{\rho}_{\mathrm{\res}}$ is the steady state of a \res{} weakly coupled to two thermal baths and a \qd{}. Moreover, the Hamiltonian of the \res{} has no degeneracies. In this situation, the standard intuition is that steady-state coherences are weak, generically of second order in the system--bath coupling strength \cite{BP, Guarnieri_2018, Trushechkin_2022, Gerry_2023} and strong \qd--\res~couplings~\cite{Sevitz2025}. This suggests that
\begin{align}
    \alpha_c = - \lambda \av{\hat{n}} + O(\Gamma_\nu^2),
\end{align}
in all coupling regimes. Although the argument above is not rigorous, this relation is confirmed by our numerical simulations.

\begin{figure}[htbp]
  \centering
  \includegraphics[width=\linewidth]{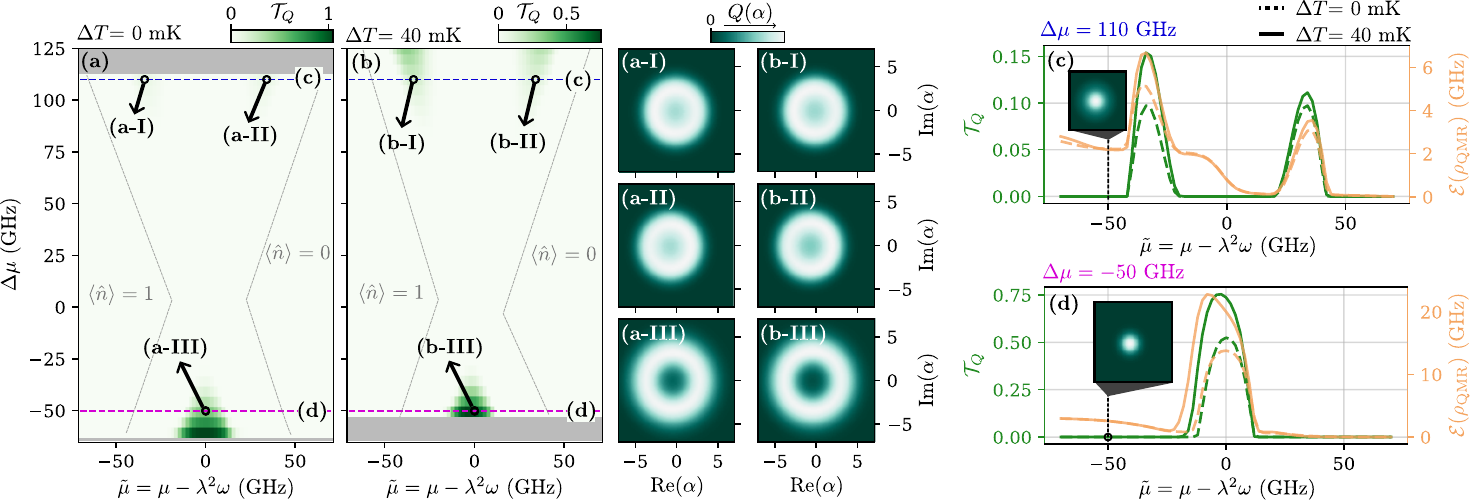}
  \caption{
  Torotropy $\TT_Q$ as a function of the dot re-normalized energy $\tilde{\mu}$ and the chemical potential imbalance $\Delta \mu$ with $T_L=T_R=13.09$ GHz ($100$ mK) in panel \textbf{(a)} and $T_L=13.09$ GHz ($100$ mK) and $T_R=7.86$ GHz ($60$ mK) in panel \textbf{(b)}. The gray regions mask the areas where data points where not computed due to high computation resources. 
  Panels \textbf{(a-I)}-\textbf{(a-III)} are the Husimi distributions corresponding to the corresponding marks in panel \textbf{(a)} while panels \textbf{(b-I)}-\textbf{(b-III)} corresponding to marks in panel \textbf{(b)}. 
  Panels \textbf{(c)} and \textbf{(d)} correspond to the line cuts of panels \textbf{(a)}-\textbf{(b)} at $\Delta\mu=-50$ GHz (blue) and $\Delta\mu=100$ GHz (magenta) respectively. Dashed lines corresponds to $\Delta T=0$ mK (panel \textbf{(a)}) while solid line is $\Delta T=40$ mK (panel \textbf{(b)}). Green lines denotes the \dnt{} and orange lines is the ergotropy of the \res~in the \NESS.
  Parameters: $\Gamma_L/(2\pi)=\Gamma_R/(2\pi)=0.2$ GHz, $\omega/(2\pi)=1$ GHz, $\lambda=0.7$, $\delta_L=\delta_R=10$ GHz, and $\gamma_L=-\gamma_R=-10$ GHz.
  }
  \label{fig:Fig6_appendix}
\end{figure}

\subsection{Ergotropy vs \dnt}
\label{app:ergo_vs_toro}

Here we give further details on the \dnt{} defined in Eq.~\eqref{eq:doughnutropy} of the main text. We introduced this measure to be able to reliably distinguish \mm{} from heating and quantify the quality of \mm. In Sec.~\ref{Sec:EmWit_SO} we studied the case when the temperature gradient between the reservoirs are zero, i.e., $\Delta T=0$ mK, and thus the transport is solely due to the chemical imbalance $\Delta\mu$. A thorough study of the electron current and the effect due to \mm~in the \res~was provided in Fig.~\ref{fig:Fig2} of the main text. Here we complement the study and showcase, in Fig.~\subfigref{fig:Fig6_appendix}{a}, the \dnt{} $\TT_Q$ as a function of the re-normalized energy of the \qd{} $\tilde{\mu}$ and the chemical imbalance $\Delta \mu$. Note that $\TT_Q\neq 0$ only when the resonator presents \mm, as discussed in Sec.~\ref{Sec:EmWit_SO}. In Fig.~\subfigref{fig:Fig6_appendix}{a-I--a-III} we showcase the Husimi distributions of the parameters marked in Fig.~\subfigref{fig:Fig6_appendix}{a}. From here, it is straightforward to see that the better quality of \mm{}, the larger $\TT_Q$. For completeness, in Fig.~\subfigref{fig:Fig6_appendix}{b}, we showcase the \dnt{} when the temperature gradient between the reservoirs is as in Sec.~\ref{Sec:MachineOperation}; namely, $\Delta T = T_L-T_R = 40$ mK. Overall, regions and behavior of $\TT_Q$, and in turn \mm, are similar to the $\Delta T =0$ mK case. 

As per the standard definition \cite{Allahverdyan_2004}, the ergotropy of the \res{} is
\begin{equation}
     \mathcal{E}(\rho_{\mathrm{\res}})= \Tr(\rho_{\mathrm{\res}} H_{\mathrm{\res}})-\min_{U}  \Tr( U \rho_{\mathrm{\res}} \, U^\dagger H_{\mathrm{\res}}),
\end{equation}
with $H_{\mathrm{\res}}=\omega\bb\b$ and $U$ a unitary operation. In Ref.~\cite{Culhane2022}, it was proposed as a witness for \mm. As stated in Sec.~\ref{Sec:EmWit_SO}, our fully quantum analysis also shows that $\mathcal{E}(\rho_{\mathrm{\res}})$ cannot detect \mm{} faithfully, which is unsurprising as nonzero ergotropy may arise due to factors other than \mm. Here, we exhibit an example of this in Fig.~\subfigref{fig:Fig6_appendix}{c}, where $\Delta\mu=110$ GHz, and Fig~\subfigref{fig:Fig6_appendix}{d}, where $\Delta\mu=-50$ GHz. On both panels, we see that, for example at $\tilde{\mu}=-50$ GHz, $\mathcal{E}(\rho_{\mathrm{\res}}) > 0$ while there is no indication of \mm{} in the Husimi distributions. On the contrary, \dnt{} (the green line in both panels) detects \mm{} faithfully. Indeed, there are no \mm{} whenever $\TT_Q = 0$, and we observe \mm{} whenever $\TT_Q > 0$.

Finally, to showcase the deeper reason behind our definition of \dnt{}, let us consider a quantum system in a state that is diagonal in the energy eigenbasis, where the energies are ordered increasingly. In this case, the populations can be interpreted as an energy distribution function. Now, the population reordering that yields the ergotropy corresponds to the non-increasing rearrangement of this distribution. The ergotropy is then given by an integral structurally identical to the one in Eq.~\eqref{eq:doughnutropy}. In other words, barring the entropic term, ergotropy is the \dnt{} of the energy distribution. Note that there is in fact no need to adhere to diagonal states---one may instead consider the so-called incoherent ergotropy \cite{Francica_2020} and allow for arbitrary states.

\section{Computation of machine thermodynamics}\label{Appendix:Engine_thermo}

To deduce the machine operation we are interested in the heat and work exchange between the system composed of the \qd+\res~and the leads. For this we first look at the energy conservation 
\begin{equation}
    0=\partial_t\langle \hat{H}\rangle \sim \partial_t \langle \hat{H}_s\rangle + \partial_t \langle \hat{H}_\R\rangle
\end{equation}
where $\hat{H}_s$ and $\hat{H}_\R$ are the system and leads Hamiltonians defined in Eq.~\eqref{eq:H_s} and Eq.~\eqref{eq:H_R} respectively. Furthermore, we made use of that we restricted our analysis to the weak systems-leads coupling such that the energy stores in the coupling $\hat{V}$ is neglected~\cite{Potts2024}. Now we focus on the change in the energy of the system is given by
\begin{align}
  \partial_t\langle \hat{H}_s\rangle = \Tr_s\left( \hat{H}_s \, \partial_t \rho_s\right) = \Tr_s\left( \tilde{H}_s \,  \partial_t \trho_s\right)
  = \Tr\left( (\omega \bb\b+\tilde{\mu} \d^\dagger \d) \,  \partial_t \trho_s \right) 
\end{align}
where we used the that the polaron transformation defined in Sec.~\ref{Sec:MachineOperation} is unitary and made use of Eq.~\eqref{eq:H_s_tilde}. Next we re-arrange the terms in the following way
\begin{align}
  \partial_t\langle \hat{H}_s\rangle = \omega\Tr\left( \bb\b \,  \partial_t \trho_s \right) + \tilde{\mu} \Tr\left(\d^\dagger \d \, \partial_t \trho_s \right) = \omega\Tr\left( \bb\b \, \partial_t \trho_s \right) - \tilde{\mu} \sum_\nu \I , 
\end{align}
where we have used total particle conservation, i.e. $\partial_t \av{\d^\dagger \d} =-\sum_\nu \partial_t \av{\c^\dagger \c}$, and in the last equality identified the second term as the electronic particle current defined in Eq.~\eqref{def:particle_current} of the main text. At this point we follow the typical approach to obtain the ``heat'' and ``work'' expressions from a \qd, where one sums and subtracts the term $\sum_\nu \mu_\nu \I$ resulting in 
\begin{align}\label{Eq_app:Hs_before_names}
  \partial_t\langle \hat{H}_s\rangle  = \omega\Tr\left( \bb\b \,  \partial_t \trho_s \right) + \omega \lambda^2 \sum_\nu \I - \sum_\nu (\mu-\mu_\nu) \I- \sum_\nu \mu_\nu \I.
\end{align}
From here we can apply the first law of thermodynamics and separate the electrical power defined as
\begin{equation}
    \Power=\sum_\nu \mu_\nu \I
\end{equation}
from the typical electrical heat currents
\begin{equation}
    \J^\mathrm{el} = (\mu-\mu_\nu) \I.
\end{equation}
Everything that is remaining in Eq.~\eqref{Eq_app:Hs_before_names} we identify as the mechanical heat currents, i.e., the energy that is injected to (or from) the \res~each time and electron jumps form (or to) lead $\nu$. The next step consist in identifying with more clarity each contribution. For this, we make use of the master equation detailed in Eq.~\eqref{Eq_app:ME_Redfield_final} as
\begin{align}\nonumber
    \Tr\left( \bb\b \,  \partial_t \trho_s \right) &= \sum_{j}\sum_{n} j\langle j|  \partial_t \trho_s^n |j\rangle 
    = \sum_\nu  \sum_{jkl}\sum_{n} j \Big( -\Rnn{n,}{n}{j}{j}{k}{l} \bra{k} \trho_s^n \ket{l} + \Rnn{n+1,}{n}{j}{j}{k}{l} \bra{k} \trho_s^{n+1} \ket{l} + \Rnn{n-1,}{n}{j}{j}{k}{l} \bra{k} \trho_s^{n-1} \ket{l} \Big) \\
    &= \sum_\nu  \sum_{jkl} j \Big( \left(\Rnn{0}{1}{j}{j}{k}{l} -\Rnn{0}{0}{j}{j}{k}{l}  \right) \bra{k} \trho_s^0 \ket{l} - \left( \Rnn{1}{1}{j}{j}{k}{l} - \Rnn{1}{0}{j}{j}{k}{l} \right) \bra{k} \trho_s^1 \ket{l} \Big).
\end{align}
From here we define generalized Mechanical tensors as 
\begin{align}\label{Eq_app:Q_01}
  \Qnn{0}{1}{j}{m}{k}{l} &= \Rnn{0}{1}{j}{m}{k}{l} - \Rnn{0}{0}{j}{m}{k}{l} 
   \\ \label{Eq_app:Q_10}
  \Qnn{1}{0}{j}{m}{k}{l} &= \Rnn{1}{1}{j}{m}{k}{l}- \Rnn{1}{0}{j}{m}{k}{l}, 
\end{align}
thus arrive to 
\begin{align}\label{Eq_app:bb_b_average}
    \Tr\left( \bb\b \, \partial_t \trho_s \right) = \sum_\nu  \sum_{jkl} j \Big( \Qnn{0}{1}{j}{m}{k}{l} \bra{k} \trho_s^0 \ket{l} - \Qnn{1}{0}{j}{m}{k}{l} \bra{k} \trho_s^1 \ket{l} \Big).
\end{align}
Replacing Eq.~\eqref{Eq_app:bb_b_average} into Eq.~\eqref{Eq_app:Hs_before_names} we obtain
\begin{align}\label{Eq_app:Hs_after_names}
  \partial_t\langle \hat{H}_s\rangle  = \sum_\nu \Bigg(  \sum_{jkl} \omega j \Big(\Qnn{0}{1}{j}{m}{k}{l} \bra{k} \trho_s^0 \ket{l} - \Qnn{1}{0}{j}{m}{k}{l} \bra{k} \trho_s^1 \ket{l} \Big) + \omega \lambda^2 \I \Bigg) -  \sum_\nu \J^\mathrm{el}  - \tilde{P}^\mathrm{el}_\mathrm{tot}.
\end{align}
where we identify the mechanical heat current arising from lead $\nu$ as
\begin{equation}
    \J^\mathrm{mec}= - \Bigg( \sum_{jkl} \omega j \Big(\Qnn{0}{1}{j}{m}{k}{l} \bra{k} \trho_s^0 \ket{l} - \Qnn{1}{0}{j}{m}{k}{l} \bra{k} \trho_s^1 \ket{l} \Big) + \omega \lambda^2 \I \Bigg),
\end{equation}
where the global minus in the front is to follow the same sign convention as the other currents where the currents are positive if the particle flows towards the reservoir $\nu$, and negative otherwise. With this we conclude the section.

\section{Details of the mechanical heat currents}\label{Appendix:HeatCurrents}

\begin{figure}[htbp]
  \centering
  \includegraphics[width=.8\linewidth]{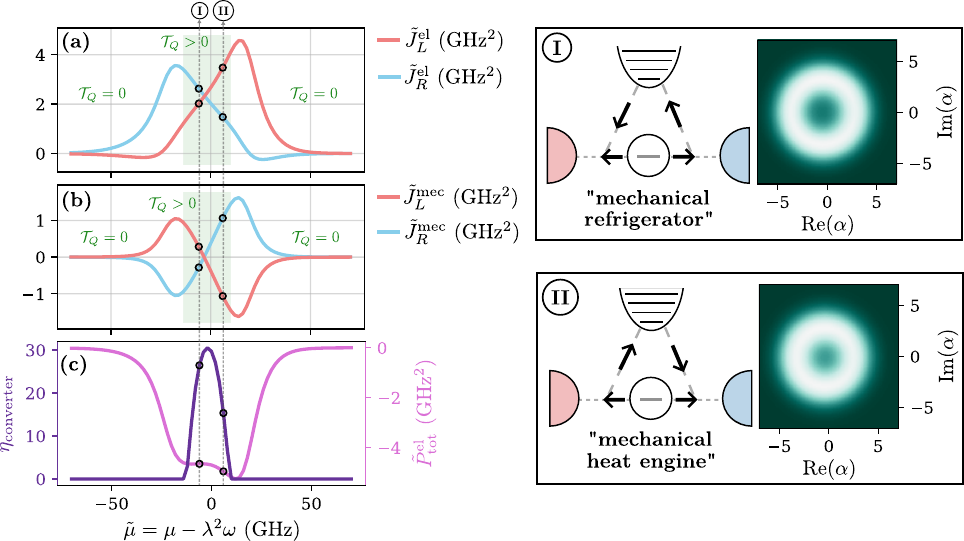}
  \caption{
  \textbf{(a)} Electrical heat currents $\JL^\mathrm{el}$ (red) and $\JR^\mathrm{el}$ (blue) as a function of the dot's re-normalization energy $\tilde{\mu}$.
  \textbf{(b)} Mechanical heat currents $\JL^\mathrm{mec}$ (red) and $\JR^\mathrm{mec}$ (blue) as a function of the dot's re-normalization energy $\tilde{\mu}$. 
  Both in panel \textbf{(a)} and \textbf{(b)} the regions where $\TT_Q>0$ indicating where \mm~occurs are highlighted in green. 
  \textbf{(c)} Conversion performance $\eta_\mathrm{converter}$ (purple) and total power $\tilde{P}^\mathrm{el}_\mathrm{tot}$ (magenta) as a function of the dot's re-normalization energy $\tilde{\mu}$. 
  Two dashed lines run across the panels \textbf{(a)}-\textbf{(c)} at $\tilde{\mu}=-6$ GHz to indicate case \textbf{I} and at $\tilde{\mu}= 6$ GHz for case \textbf{II}.
  Parameters: $\Gamma_L/(2\pi)=\Gamma_R/(2\pi)=0.2$ GHz, $\omega/(2\pi)=1$ GHz, $T_L=13.09$ GHz ($100$ mK), $T_R=7.86$ GHz ($60$ mK) and $\delta_L=\delta_R=10$ GHz, $\delta_L=\delta_R=10$ GHz, $\gamma_L=-\gamma_R=-10$ GHz, $\lambda=0.7$, and $\Delta\mu -45 $ GHz. 
  }
  \label{fig:Fig7_appendix}
\end{figure}

This section is to complement the previous section and Sec.~\ref{Sec:MachineOperation} of the main text where we derived and exposed the electrical and mechanical heat currents. Here we follow the same sign convention where currents are positive if the particle flows towards the reservoir $\nu$ and negative otherwise.

On the one hand, the electrical heat currents $\JL^\mathrm{el}$/$\JR^\mathrm{el}$ (defined in Eq.~\eqref{eq:J_el} of the main text) quantifies the energy the electron transports from the reservoir $\nu$ to the \qd~under the polaron frame. On the other hand the mechanical heat currents $\JL^\mathrm{mec}$/$\JR^\mathrm{mec}$ (defined in Eq.~\eqref{eq:J_mec} of the main text) measures the energy allocated to (from) the \res~due to the electronic transport from the reservoir $\nu$ to the \qd. In the main text we state that these currents are strictly zero when the \qd--\res~coupling is zero. To see this, we first explicitly write down the Redfield tensors (defined in Eq.~\eqref{Eq_app:A00}-\eqref{Eq_app:A10} of the main text) with $\lambda=0$~\cite{Sevitz2025}
\begin{align}
  \Rnn{1}{1}{j}{m}{k}{l} & = \delta_{j,k} \delta_{ml} R^{1\rightarrow 0}_\nu(\mu) , 
  & \Rnn{0}{0}{j}{m}{k}{l} & =  \delta_{j,k} \delta_{ml} R^{0\rightarrow 1}_\nu(\mu) , \\ 
  \Rnn{1}{0}{j}{m}{k}{l} & = \delta_{j,k}\delta_{l,m} R^{1\rightarrow 0}_\nu(\mu) , 
  & \Rnn{0}{1}{j}{m}{k}{l} & = \delta_{j,k}\delta_{l,m} R^{0\rightarrow 1}_\nu(\mu). 
\end{align}
Replacing these expressions into the Mechanical tensors defined in Eq.~\eqref{Eq_app:Q_01}-\eqref{Eq_app:Q_10} yields
\begin{align}
  \Qnn{0}{1}{j}{m}{k}{l} &= \Rnn{0}{1}{j}{m}
  {k}{l} - \Rnn{0}{0}{j}{m}{k}{l} = \delta_{j,k}\delta_{l,m} \Big( R^{0\rightarrow 1}_\nu(\mu) - R^{0\rightarrow 1}_\nu(\mu) \Big) = 0, \\ 
  \Qnn{1}{0}{j}{m}{k}{l} &= \Rnn{1}{1}{j}{m}{k}{l}- \Rnn{1}{0}{j}{m}{k}{l} = \delta_{j,k} \delta_{ml} \Big( R^{1\rightarrow 0}_\nu(\mu) - R^{1\rightarrow 0}_\nu(\mu) \Big) = 0 . 
\end{align}
Thus with this it is straightforward to observe that $\JL^\mathrm{mec}=\JR^\mathrm{mec}=0$ if $\lambda=0$, providing the evidence needed for our claim.

Returning to the case when the \qd--\res~coupling exists $\lambda \neq 0$, we now study in detail their behavior to complement the analysis and results of Sec.~\ref{Sec:MachineOperation}. We plot the electrical heat currents in Fig.~\subfigref{fig:Fig7_appendix}{a} and the mechanical heat currents in Fig.~\subfigref{fig:Fig7_appendix}{b} as a function of the dot's re-normalization energy $\tilde{\mu}$. For the sake of the example, we have chosen a situation where \mm~occur corresponding to a fixed $\Delta\mu -45 $ GHz. In green we highlight the values of $\tilde{\mu}$ for which \mm~is enabled, dictated by the \dnt{} $\TT_Q$. In this region both of the electrical heat currents ($\JL^\mathrm{el}$ and $\JR^\mathrm{el}$) are positive. Now, if we look at the mechanical heat currents, \mm~are enabled under two situations: $\JL^\mathrm{mec}>0$ and $\JR^\mathrm{mec}<0$ which we call case \textbf{I}, and $\JL^\mathrm{mec}<0$ and $\JR^\mathrm{mec}>0$ which we refer to as case \textbf{II}. For clarity in our discussion, in the panels to the right we explicitly indicate with arrows the direction of the current flow. In case \textbf{I}, the electrical currents heats both reservoirs while the mechanical component absorbs heat from the cold bath and dissipates onto the hot bath. This resembles the action of a refrigerator, but not exactly as there is no mechanical power input. For this reason we define this operation as a ``mechanical refrigerator'' in quotation marks. On the contrary, in case \textbf{II}, the electrical currents again heats both reservoirs but now the mechanical component absorbs heat from the cold bath and dissipates onto the hot bath. Exactly the opposite as in case \textbf{I}. Here the operation resembles that of a heat engine, thus the name ``mechanical heat engine''. 

An interesting questions at this point is: which of the two disconsider (case \textbf{I} or \textbf{II}) is most favorable for the autonomous machine that converts PE to \mm? To answer this question in Fig.~\subfigref{fig:Fig7_appendix}{c} we plot the conversion performance measure defined in Eq.~\eqref{eq:Performance_converter} of the main text. The performance of case \textbf{I} is larger than in case \textbf{II}. It should be noted that the power $\tilde{P}^\mathrm{el}_\mathrm{tot}$ is larger for case \textbf{I} than case \textbf{II}. Thus an immediate answer to the question would be case \textbf{I}. Though, the consequence and underlining understanding of these results is still not clear to the authors, thus further work has to be carried out.

\section{Transport properties for temperature bias}\label{Appendix:TransportProp_Delta_T_40}

\begin{figure}[htbp]
  \centering
  \includegraphics[width=\linewidth]{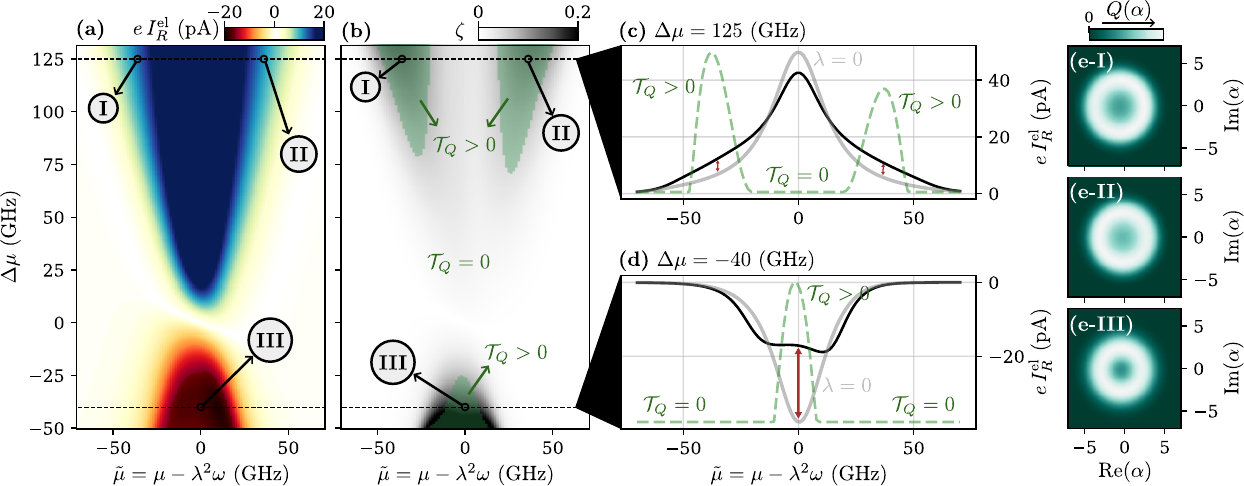}
  \caption{
  \textbf{(a)} $\IR$ as a function of both $\tilde{\mu}$ and $\Delta\mu$. Blue values correspond to left to right electronic hopping ($\IR>0$) and red values to the inverse direction ($\IR<0$). Due to the temperature imbalance among the reservoirs, the sharp changes in the current that witness \mm~in the \res~are not as clear as in the case where $\Delta T =0$ mk. 
  \textbf{(b)} The \mm{} witnessing parameter $\zeta$, given in Eq.~\eqref{eq:SO_condition_R} of the main text, as a function of both $\tilde{\mu}$ and $\Delta\mu$. On-top we have delimited in green the region where \mm~are present following the \dnt{} measure.
  In panels \textbf{(c)} and \textbf{(d)} we plot a line cut of panel \textbf{(a)} at $\Delta\mu=125$ GHz and $\Delta\mu=-40$ GHz respectively. 
  To observe the sharp change of monotonicity on the currents, we have included the particle current for the no-coupling $\lambda=0$ situation in gray. With red arrows me highlight the regions where there is a sharp change in the behavior fo the current which corresponds to the regions where $\TT_Q>0$ (plotted in dashed green), that is where \mm~in the \res~are present.
  To evidence the state of the resonator, we compute Husimi distribution plots for $\Delta\mu = 125$ GHz and $\tilde{\mu}=-36$ GHz in panel \textbf{(e-I)}, $\Delta\mu =125$ GHz and $\tilde{\mu}=36$ GHz in panel \textbf{(e-II)} and $\Delta\mu=-40$ GHz and and $\tilde{\mu}=0$ GHz in panel \textbf{(e-III)}.
  Parameters: $\Gamma_L/(2\pi)=\Gamma_R/(2\pi)=0.2$ GHz, $\omega/(2\pi)=1$ GHz, $T_L= 13.09$ GHz ($100$ mK), $T_R=7.86$ GHz ($60$ mK), $\lambda=0.7$, $\delta_L=\delta_R=10$ GHz and $\gamma_L=-\gamma_R=-10$ GHz.
  }
  \label{fig:Fig8_appendix}
\end{figure}

In this section we showcase the transport properties for the parameters corresponding to Fig.~\ref{fig:Fig3} in Sec.~\ref{Sec:MachineOperation}. In Fig.~\subfigref{fig:Fig8_appendix}{a} we plot the $\IR$ as a function of both $\tilde{\mu}$ and $\Delta\mu$ for a constant \qd--\res~coupling of $\lambda=0.7$. Note that unlike the $\Delta T=0$ mK case, for $\Delta \mu=0$ GHz $\IR\neq 0$, this this is due to the thermal imbalance among the baths. For a finer detail of the presence of thermo currents, we refer the reader to Ref.~\cite{Sevitz2025}. In Fig.~\subfigref{fig:Fig8_appendix}{b} we observe the regions where \mm~are present dictated by the \dnt{} $\TT_Q$ (see panels \textbf{(e-I)}-\textbf{(e-III)} for the Husimi distributions). Following our discussion in Sec.~\ref{Sec:EmWit_SO} we observe that the regions where $\TT_Q>0$ aligned with the regions where $\zeta$ is large.

\section{Machine operation for different couplings}

\begin{figure}[htbp]
  \centering
  \includegraphics[width=\linewidth]{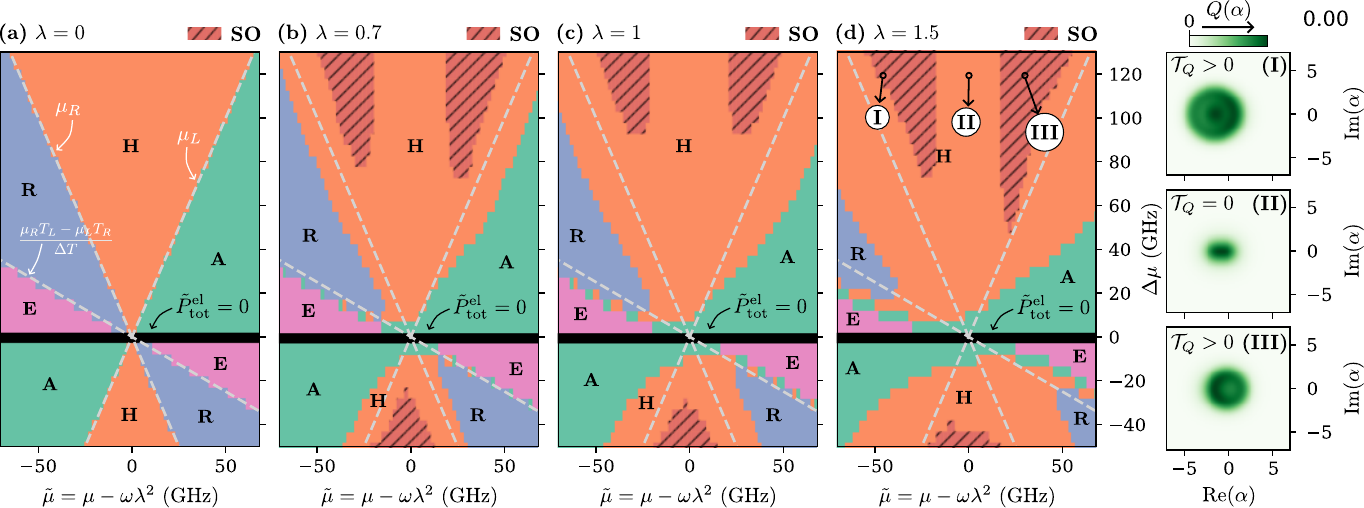}
  \caption{Complementary operation modes of the autonomous machine for couplings corresponding to: \textbf{(a)} $\lambda=0$, \textbf{(b)} $\lambda=0.7$, \textbf{(c)} $\lambda=1$, and \textbf{(d)} $\lambda=1.5$. The color coding is the following; pink Heat Engine (\textbf{HE}), blue Refrigeration (\textbf{R}), orange Heater (\textbf{P}) and with green Accelerator (\textbf{A}). The regions where \mm~(\textbf{SO}) are present are denoted in red with diagonal black lines. In panel \textbf{(d)} we highlight three points with $\Delta\mu=120$ GHz and plot their respective Husimi distributions in panels \textbf{(I)} with $\tilde{\mu}=-46$ GHz, \textbf{(II)} with $\tilde{\mu}=0$ GHz, and \textbf{(III)} with $\tilde{\mu}=30$ GHz. 
  Parameters: $\Gamma_L/(2\pi)=\Gamma_R/(2\pi)= 0.2$ GHz, $\omega/(2\pi)=1$ GHz, $T_L= 13.09$ GHz ($100$ mK), $T_R=7.86$ GHz ($60$ mK) and $\delta_L=\delta_R=10$ GHz and $\gamma_L=-\gamma_R=-10$ GHz.
  }
  \label{fig:Fig9_appendix}
\end{figure}

In Sec.~\ref{Sec:MachineOperation} of the main text we analyzed the effect on the \qd--\res~coupling over the operation of the PE machine. In Fig.~\subfigref{fig:Fig1}{c-d} we showcased the machine operation for couplings $\lambda=0$ and $\lambda=0.7$ respectively. Here we complement that study and show the effect for larger couplings. In Fig.~\ref{fig:Fig9_appendix} we exhibit the machine operation for couplings \textbf{(a)} $\lambda=0$, \textbf{(b)} $\lambda=0.7$, and \textbf{(c)} $\lambda=1$. We note that the higher the \qd--\res~coupling the more the heater regime seeps onto the neighboring regions of operation. Furthermore, regardless of the coupling, the region where \mm~is present always corresponds to the heater operation.

Although in the main text we focused on the cases where $\lambda\leq1$, here we also include the operation of \qd--\res~coupling $\lambda=1.5$ in Fig.~\subfigref{fig:Fig9_appendix}{d}. The behavior of the operation is similar as in the $\lambda\leq1$ cases and \mm~is only present in the heater operation. We also plot the Husimi distribution for three representative cases in panels labeled (I)-(III). As anticipated the distributions are asymmetric which have been as extensively reported for other platforms in~\cite{Lim2024,Dutta2019,Ben2021,Lee2013}.

\section{Thermodynamics performance measure}

\begin{figure}[htbp]
  \centering
  \includegraphics[width=0.5\linewidth]{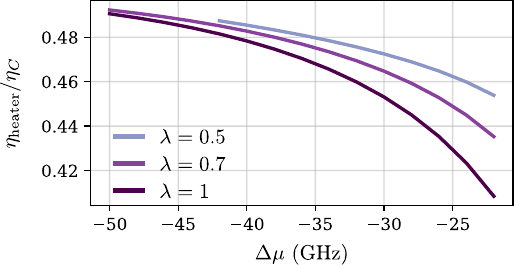}
  \caption{Heater performance $\eta_\mathrm{heater}$ normalized with the Carnot efficiency $\eta_C$ as a function of the chemical imbalance $\Delta\mu$ among the reservoir with a fixed re-normalized energy $\tilde{\mu}=0$ GHz for \qd~+\res~couplings $\lambda=0.5$, $\lambda=0.7$, and $\lambda=1$.
  Parameters: $\Gamma_L/(2\pi)=\Gamma_R/(2\pi)=0.2$ GHz, $\omega/(2\pi)=1$ GHz, $T_L= 13.09$ GHz ($100$ mK), $T_R=7.86$ GHz ($60$ mK) and $\delta_L=\delta_R=10$ GHz, $\delta_L=\delta_R=10$ GHz and $\gamma_L=-\gamma_R=-10$ GHz.}
  \label{fig:Fig10_appendix}
\end{figure}

In Sec.~\ref{Sec:Performance} of the main text we quantified how ``well'' the autonomous quantum machine converts PE to \mm~by defining the performance measure $\eta_\mathrm{converter}$ (see Eq.~\eqref{eq:Performance_converter}). Here, instead, we compute the thermodynamics heater performance given in Ref.~\cite{Manzano2020} as
\begin{equation}\label{Eq_app:eta_pump}
  \eta_\mathrm{heater}= \frac{|\JL^\mathrm{tot}|(1-T_\mathrm{ref}/T_L)}{|\JR^\mathrm{tot}|(1-T_\mathrm{ref}/T_R)+|\Power|} \leq 1,
\end{equation}
where $T_\mathrm{ref}$ is a reference temperature such that $T_L\leq T_\mathrm{ref} \leq T_R$. This quantity measures how well the power allocated from the chemical imbalance reveres the flow of heat form the hot reservoir. In the limit of $T_\mathrm{ref} \rightarrow T_R$ Eq.~\eqref{Eq_app:eta_pump} reduces to
\begin{equation}\label{Eq_app:eta_pump_limit}
  \lim_{T_\mathrm{ref} \rightarrow T_R}\eta_\mathrm{heater}= \eta_C\frac{|\JL^\mathrm{tot}|}{|\Power|}
\end{equation}
with $\eta_C=1-T_R/T_L$ the typical Carnot efficiency. In Fig.~\ref{fig:Fig10_appendix} we plot Eq.~\eqref{Eq_app:eta_pump_limit} function of the chemical imbalance $\Delta \mu$ for different \qd--\res~coupling. The parameters are the same as analyzed in Sec.~\ref{Sec:Performance}. 

Similar to the the behavior of the machine conversion performance defined in Eq.~\eqref{eq:Performance_converter} of the main text: for a specific value of $\Delta \mu$ the lower \qd--\res~coupling, the higher $\eta_\mathrm{heater}$ is. Furthermore, comparing with Fig.~\ref{fig:Fig4} of the main text we observe that the regions where \mm~are present, $\eta_\mathrm{heater}$ plateaus to a constant value. This leads us to believe there might be an underlining thermodynamic relationship between Eq.~\eqref{eq:Performance_converter} and Eqs.~\eqref{Eq_app:eta_pump}--\eqref{Eq_app:eta_pump_limit} though more work has to be done to draw out any conclusions. 

\end{document}